\documentclass[10pt,conference,compsocconf,letterpaper]{IEEEtran}

\IEEEoverridecommandlockouts

\usepackage{amsfonts}
\usepackage{amsmath}
\usepackage{amssymb}
\usepackage{graphicx}
\usepackage[boxed,ruled,vlined,linesnumbered]{algorithm2e}
\usepackage{paralist}
\usepackage{flushend}

\usepackage{xcolor}

\ifCLASSINFOpdf
\else
\fi

\begin{document}

\title{Understanding the Timed Distributed Trace of a Partially Synchronous System at Runtime}

\author{\IEEEauthorblockN{Yiling Yang$^{1,2}$, Yu Huang$^{1,2}$\IEEEauthorrefmark{1}, Jiannong Cao$^3$, Jian Lu$^{1,2}$}
\thanks{\IEEEauthorrefmark{1}Corresponding author.}
\IEEEauthorblockA{$^1$State Key Laboratory for Novel Software Technology\\
Nanjing University, Nanjing 210046, China\\
$^2$Institute of Computer Software\\
Nanjing University, Nanjing 210046, China\\
csylyang@gmail.com, \{yuhuang, lj\}@nju.edu.cn}
$^3$Internet and Mobile Computing Lab, Department of Computing\\
Hong Kong Polytechnic University, Hong Kong, China\\
csjcao@comp.polyu.edu.hk
}

\maketitle

\begin{abstract}
    It has gained broad attention to understand the timed distributed trace of a cyber-physical system at runtime, which is often achieved by verifying properties over the observed trace of system execution. However, this verification is facing severe challenges. First, in realistic settings, the computing entities only have imperfectly synchronized clocks. A proper timing model is essential to the interpretation of the trace of system execution. Second, the specification should be able to express properties with real-time constraints despite the asynchrony, and the semantics should be interpreted over the currently-observed and continuously-growing trace. To address these challenges, we propose \textsf{PARO} - the \textit{\underline{par}tially synchronous system \underline{o}bservation} framework, which i) adopts the partially synchronous model of time, and introduces the lattice and the timed automata theories to model the trace of system execution; ii) adopts a tailored subset of TCTL to specify temporal properties, and defines the 3-valued semantics to interpret the properties over the currently-observed finite trace; iii) constructs the timed automaton corresponding to the trace at runtime, and reduces the satisfaction of the 3-valued semantics over finite traces to that of the classical boolean semantics over infinite traces. \textsf{PARO} is implemented over MIPA - the open-source middleware we developed. Performance measurements show the cost-effectiveness of \textsf{PARO} in different settings of key environmental factors.
\end{abstract}

\IEEEpeerreviewmaketitle

\section{Introduction} \label{sec:Introduction}

Advances in sensor and actuator technologies have given rise to the influx of an increasing number of mobile robots with sensing and controlling abilities, besides the basic abilities of computation and communication. Distributed systems formed by the interconnections of such mobile robots enable a variety of novel applications. A typical example of such distributed systems is a group of mobile robots for collaborative tasks, e.g., patrolling a chemical plant \cite{Zhan13, Duggirala12, Yang15}.

Though the systems of collaborating mobile robots enable various novel applications, they are notoriously difficult to program. Each mobile robot executes a program implementing one or more distributed algorithms, moves and manipulates its environment, and exchanges messages over wireless communication channels with other robots. Even if the high-level distributed algorithms for these systems are well-understood, failures, timing-errors, and message delays make their implementation challenging. Moreover, the system of mobile robots must be aware of and adaptive to the situation they operate in. For example, the robots must know the relative locations of robots nearby, to collaboratively form some geometric pattern.

The challenges above motivate us to observe and analyze the execution of a distributed system of mobile robots at runtime. By understanding the system execution, the system developer can delineate and detect symptoms of software bugs. The system itself can also achieve awareness of its situation to adapt its behavior accordingly. Runtime observation and analysis of system execution is an important technique to improve system accountability, and is a complement to the traditional design time techniques such as model checking \cite{Baier08}.

Runtime observation and analysis of system execution is achieved by runtime verifying specified properties over the system execution trace. For example, the robots are designed to satisfy the property $C_1$: \textit{all the robots should gather at the assembly point within 15 seconds}. By runtime verifying the property over the collected execution trace of the robots, we can know whether the property is satisfied or violated. The results of the verification can further provide useful guidance to the developer (e.g., for the debugging of the system) or the system itself (e.g., for triggering the runtime adaptation of the system). However, runtime verification over the system execution trace is facing severe challenges.

The primary challenge is the intrinsic asynchrony of the system. The robots do not share the same notion of time and communications among them suffer from uncertain delay \cite{Corbett12, Duggirala12, Stoller00, Kshemkalyani13}. The challenge of the asynchrony becomes more severe due to the resource constraints on the mobile robots, the unreliability of the wireless communications, and the interaction with the physical environment. To cope with this challenge, a proper timing model fitting the actual system is essential to modeling the observed trace of the system execution.

The \textit{synchronous model} can shield the underlying asynchrony of the system and provide a total-order illusion of all the events in the trace. However, the synchronous model is overly-optimistic in that the local clocks of the robots are never perfectly synchronized in realistic settings \cite{Duggirala12, Corbett12}. Reasoning properties directly over the trace assumed to be perfectly timestamped may lead to the neglect of the potential violation of the properties \cite{Kshemkalyani13}.

The \textit{asynchronous model} is the most general model since it makes no synchrony assumptions about the underlying system \cite{Gartner99}. However, the asynchronous model is overly-pessimistic in that the existing synchrony of the system (e.g., the synchronized, though not perfect, clocks) is completely abandoned. The cost of reasoning over the asynchronous model is often prohibitively high \cite{Yang13, Wei12}. Moreover, the asynchronous model can only describe the temporal happen-before relation between events in the trace. Properties with real-time constraints (e.g., ``within 15 seconds'' in $C_1$) cannot be reasoned over the asynchronous model.

Varying from the synchronous model and the asynchronous model, the \textit{partially synchronous model} can appropriately model at various levels the synchrony which reasonably exists in a realistic distributed system, such as the imperfectly synchronized clocks in our motivating scenario. Therefore, the cost of reasoning can be effectively restricted, unlike that of the asynchronous model \cite{Stoller00}. Besides, we can also reason properties with real-time constraints with the knowledge that the trace is imperfectly timestamped, although techniques dedicated for handling the uncertainty resulting from the partially synchronous model are still in demand.

Another important challenge arises in the specification of system properties of interest to the specifier (i.e., the system developer or the system itself). The specifier usually has real-time requirements, e.g., in property $C_1$. To check these requirements over the execution trace, we are thus concerned with properties with metric-time constraints. To cope with this challenge, we should provide a formal specification mechanism which can express metric-time properties. Meanwhile, the partially synchronous model has the branching-time structure due to the uncertainty resulting from the asynchrony, i.e., we can derive multiple possible executions of the partially synchronous system, besides the actually observed one. Thus, the specification should also be able to capture the notion of branching time. In addition, as the system executes, the observed trace is continuously ``growing'' to a potentially infinite size. The specification should be interpreted over the currently-observed (but still growing) finite trace.

Discussions above necessitate a systematic scheme for formal specification and runtime verification of properties with metric-time constraints over the execution trace of a partially synchronous system. Toward this objective, we propose \textsf{PARO} - the \textit{\underline{par}tially synchronous system \underline{o}bservation} framework, which consists of three essential parts:
\begin{enumerate}\setlength{\itemsep}{0pt}
  \item {\it Modeling of the Trace of System Execution.} We adopt the partially synchronous model of time. The lattice theory is employed to model the branching structure of the system execution trace, and the timed automata theory is employed to model the metric structure of the trace. Hence, we model the system execution trace as a continuously-growing timed automaton;
  \item {\it Specification of Temporal Properties.} TCTL is adopted to specify temporal properties over the execution trace. TCTL has the branching time structure to cope with the asynchrony in the trace. It can also express the metric-time properties of concern to the specifier. We employ a tailored subset of TCTL to trade certain expressiveness for the efficiency of verification. We also define the 3-valued semantics for the TCTL formulas to be interpreted over the currently-observed finite trace;
  \item {\it Verification of the Specified Property at Runtime.} We first construct the continuously-growing timed automaton corresponding to the currently-observed trace in an incremental way. Then, we reduce the satisfaction of the 3-valued semantics over the finite trace to that of the classical boolean semantics over the infinite trace.
\end{enumerate}
The \textsf{PARO} framework is implemented and evaluated over MIPA - the open-source middleware we developed \cite{MIPA, Yang15}. The performance measurements show the cost-effectiveness of \textsf{PARO} in different settings of key environmental factors. A case study of the realistic mobile robot gathering scenario is also conducted to demonstrate the effectiveness of \textsf{PARO} (as detailed in Appendix \ref{sec:Case Study}). 

The rest of this paper is organized as follows. In Section~\ref{sec:Modeling},~\ref{sec:Specification}, and~\ref{sec:Detection}, we discuss the three essential parts of the \textsf{PARO} framework. In Section~\ref{sec:Performance measurements}, we present the implementation and performance measurements. In Section~\ref{sec:Related work}, we review the existing work. Section~\ref{sec:Conclusion} concludes the paper with a brief summary and the future work.

\section{Modeling of the Trace of System Execution} \label{sec:Modeling}

Understanding the execution of a distributed system of mobile robots is achieved by verifying specified temporal properties over the trace of system execution \cite{Cooper91}. A distributed system consists of a collection of processes $P^{(1)}, P^{(2)},\cdots, P^{(n)}$. Examples of the processes include a software process manipulating a mobile robot. One checker process $P_{che}$ is in charge of collecting the execution trace of the processes and verifying the specified property. In this section, we first discuss the partially synchronous model employed to interpret the trace of system execution. Then we discuss the modeling of the branching structure and the metric structure of the trace with the lattice theory and the timed automata theory, respectively.

\subsection{The Partially Synchronous Model} \label{sec:system model}

In realistic settings, each process $P^{(k)}$ may have a local clock, and they synchronize their local clocks with an external \textit{source clock} $T$. The external clock synchronization is widely adopted and is especially useful in loosely-coupled networks \cite{Patt94}, e.g., the NTP protocol is used for external synchronization of the Internet \cite{Mills91}.

We model the processes as a partially synchronous system with approximately-synchronized real-time clocks towards the external source clock. We assume a bound $\varepsilon$ on the difference between local clocks and the source clock.\footnote{Note that our framework allows $\varepsilon$ to vary over time, and here we assume a fixed bound $\varepsilon$ for the ease of interpretation.} That is, for each event $e$ with local clock timestamp $t$, the global time (referring to the source clock) is bounded by a time interval $I(e) = [lo,hi]$ with $lo = t - \varepsilon$ and $hi = t + \varepsilon$.\footnote{We assume that the system starts at time 0 for simplicity.} Note that we assume the time intervals of the events of the same process are non-decreasing and are consistent with the process order, and there are various ways to ensure this assumption \cite{Stoller00}. In the following sections, unless explicitly stated, we use $lo$ and $hi$ to denote the lower and upper bounds of a time interval, respectively.

\subsection{Modeling the Branching Structure of Time}

We first model the branching structure of time by the lattice of consistent global states (i.e., lattice of system snapshots). Then we define the possible temporal evolutions of the system state by the active-surface-induced CGS sequences.

\subsubsection{Lattice of Consistent Global States}

\begin{figure}[tbp]
\begin{center}
  \includegraphics[width=2.5in]{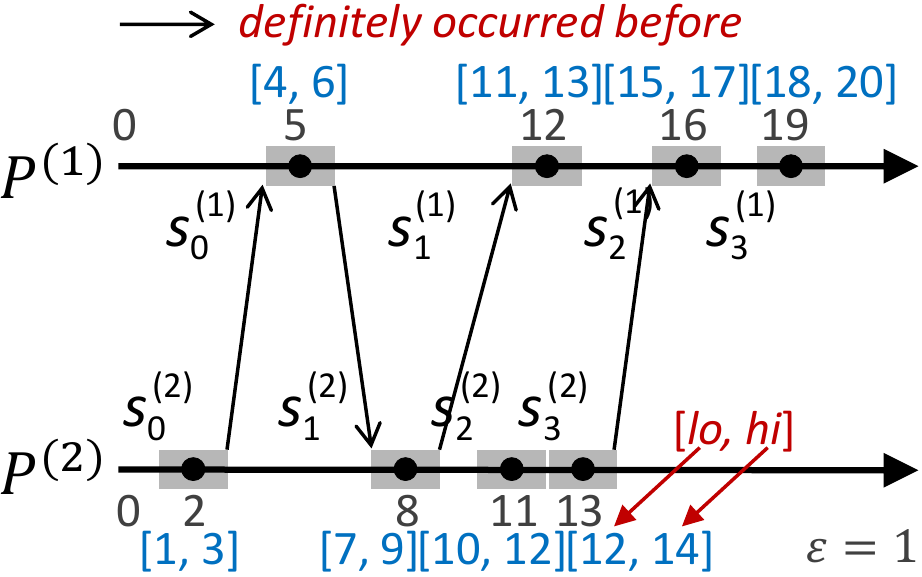}
  \caption{Space time diagram}
  \label{F:space-time diagram}
\end{center}
\end{figure}

As the system executes, each process $P^{(k)}$ generates its (potentially infinite) trace of \textit{local states} connected by \textit{events}:``$e^{(k)}_0, s^{(k)}_0, e^{(k)}_1, s^{(k)}_1, \cdots$''. Each local state $s^{(k)}_i$ is defined by the two adjacent events $e^{(k)}_i$ and $e^{(k)}_{i+1}$, denoted by $le(s^{(k)}_i)$ and $he(s^{(k)}_i)$, respectively. According to our timing model, the global time of each event can be bounded by a time interval $[lo,hi]$. Thus we can define the ``\textit{definitely occurred before}'' relation (denoted by `$\rightarrow$') between the events and the states \cite{Stoller00}. For two events $e_1$ and $e_2$, $e_1 \rightarrow e_2$ if i) they are on the same process and $e_1.lo < e_2.lo$; or ii) they are on different processes and $e_1.hi < e_2.lo$. Based on the relation between the events, we can further define the relation between local states. For two local states $s_1$ and $s_2$, $s_1 \rightarrow s_2$ if i) they are on the same process and $le(s_1) \rightarrow le(s_2)$; or ii) they are on different processes and $he(s_1) \rightarrow le(s_2)$. As shown in Fig.~\ref{F:space-time diagram}, the events (the black dots) are labeled with time intervals and the `$\rightarrow$' relation between the events is explicitly depicted.

\begin{figure}[tbp]
\begin{center}
  \includegraphics[width=2.5in]{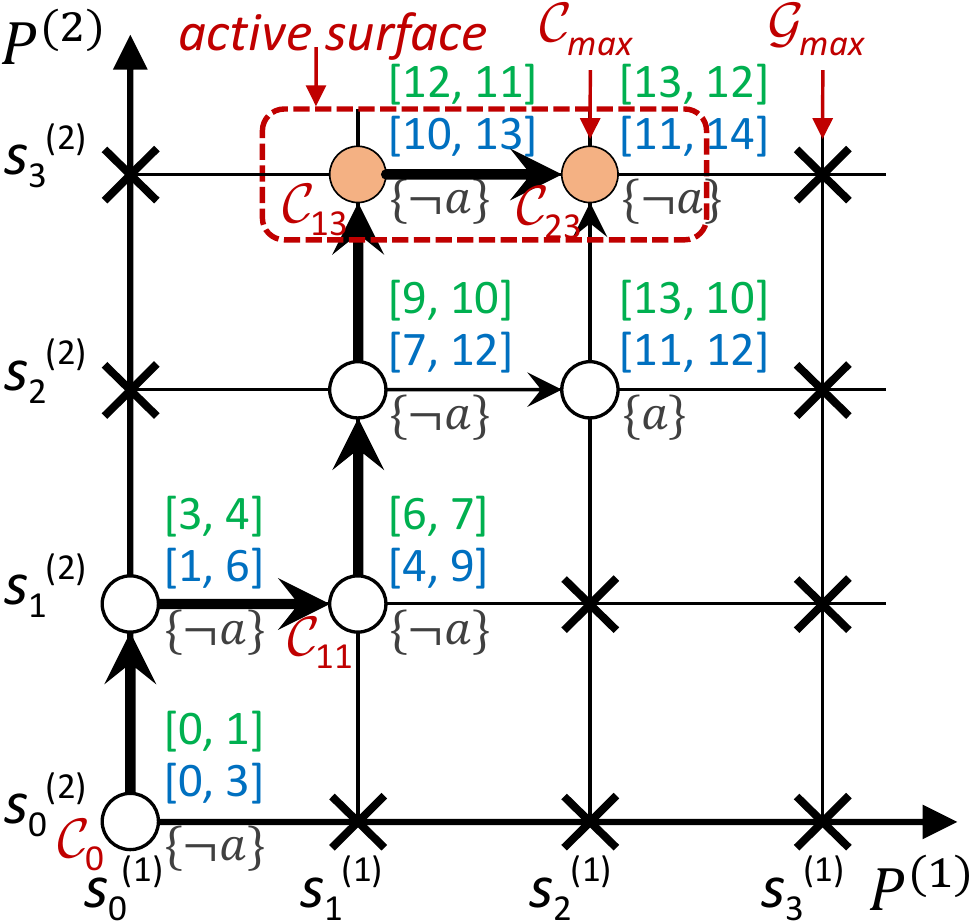}
  \caption{Lattice of CGSs}
  \label{F:Lattice}
\end{center}
\end{figure}

A \textit{global state} $\mathcal{G} = (s^{(1)}, s^{(2)}, \cdots, s^{(n)})$ is an $n$-tuple of local states from each $P^{(k)}$.
Intuitively, a global state is consistent if an omniscient external observer could possibly observe that the system enters this state. Formally, a global state $\mathcal{C}$ is consistent iff the constituent local states are pairwise concurrent, i.e.,
\begin{displaymath}
\mathcal{C} = (s^{(1)}, s^{(2)}, \cdots, s^{(n)}), \forall\ i,j : i\neq j :: \neg (s^{(i)}\rightarrow s^{(j)})
\end{displaymath}
A Consistent Global State (CGS) denotes a snapshot or a meaningful observation of the distributed system \cite{Babaoglu93, Schwarz94}. We use $\mathcal{C}[k]$ to denote the $k^{th}$ constituent local state of CGS $\mathcal{C}$.
One key notion is that the set of observed CGSs 
has the lattice structure (denoted by $LAT$) \cite{Stoller00}. The significance of time is that it restricts the possible interleavings of local states, which in turn determines the lattice of system snapshots. As the system executes, the observed lattice ``grows'' at runtime, to a potentially infinite size. Fig.~\ref{F:Lattice} shows a currently-observed lattice of Fig.~\ref{F:space-time diagram}. The dots `$\bigcirc$' denote the CGSs and crosses `$\times$' denote inconsistent global states.

We specify predicates over the CGSs to delineate properties concerning specific snapshot of the system. The predicates over CGSs can be viewed as the labeling of CGSs with letters from a finite alphabet $AP$ (i.e., all pre-defined CGS predicates). Fig.~\ref{F:Lattice} is an example where each CGS is labeled with the predicates (`$a$' or `$\neg a$') it satisfies.

\subsubsection{Active-surface-induced CGS Sequences}

As the system executes, the lattice ``grows'' and new CGSs will be added to {\it LAT} as successors of the \textit{active surface CGSs} \cite{Yang13}. CGSs whose immediate successors (consistent or inconsistent global states) are not all discovered could have new immediate successors. We define these CGSs as the \textit{active surface}. To formally define the active surface, first note that $P_{che}$ uses a queue $Que^{(k)}$($1\leq k \leq n$) to store the local states (in FIFO manner) sent from each $P^{(k)}$ \cite{Garg96, Huang12, Yang13}. Let $\mathcal{G}_{max}$ be the maximal observed global state (not necessarily consistent). The active surface is defined as:
$$Act(LAT) = \{\mathcal{C}~|~\mathcal{C}\in LAT, \exists k, \mathcal{C}[k] = \mathcal{G}_{max}[k]\}$$
We are concerned with the active surface, because when $P_{che}$ observes a new local state from some process $P^{(k)}$, new CGSs stem from the active surface \cite{Yang13}. An example of the active surface is shown in Fig.~\ref{F:Lattice}.

In order to model the temporal evolution of the system state, we define the {\it CGS sequence} $Seq(\mathcal{C}_i, \mathcal{C}_j)$ as a sequence of CGSs. The bold line in Fig.~\ref{F:Lattice} denotes a CGS sequence. We use $Seq[k]$ to denote the $(k+1)^{st}$ CGS of the CGS sequence $Seq$.
%
%
Active-surface-induced CGS sequences, i.e., CGS sequences which originate from the initial CGS and
span to the active surface CGSs, capture all possible temporal evolutions of the system state resulting from the asynchrony \cite{Yang13}. They are observed finite prefixes of the potentially infinite system state evolutions. Active-surface-induced CGS sequences of $LAT$ are defined as:
\begin{eqnarray*}
  Path(LAT) = \{Seq(\mathcal{C}_0,\mathcal{C}_i)~|~\mathcal{C}_i\in Act(LAT)\}
\end{eqnarray*}
For example, in Fig.~\ref{F:Lattice}, all possible temporal evolutions of the system state are CGS sequences starting from $\mathcal{C}_0$ and currently ending at the active surface CGSs $\{\mathcal{C}_{13}, \mathcal{C}_{23}\}$.

Please refer to our previous work for more detailed discussions on the modeling of asynchronous computations \cite{Yang13, Huang12, Wei12, Yang14}.

\subsection{Modeling the Metric Structure of Time}

In this section, we discuss the modeling of the metric structure of time, based on the lattice structure discussed above. We first discuss the modeling of time information of the CGSs of the lattice, and then discuss the modeling of the lattice as a timed automaton.

\subsubsection{Time Information of the CGSs}
\begin{figure}[tbp]
\begin{center}
  \includegraphics[width=2.5in]{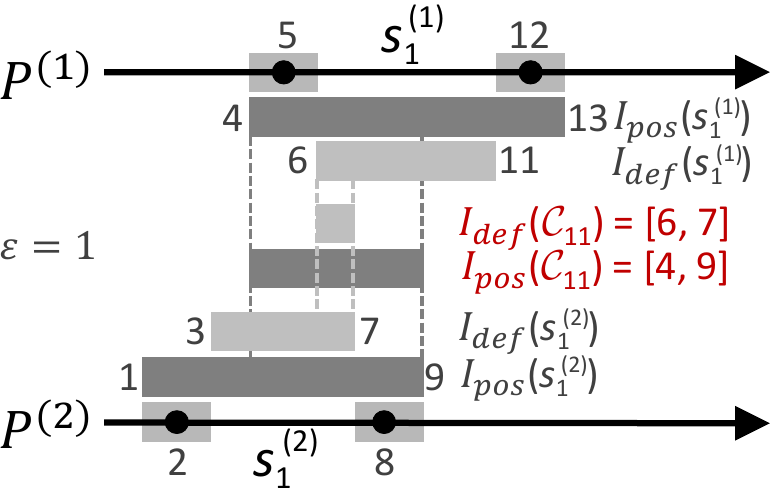}
  \caption{Definite and possible intervals of a CGS}
  \label{F:di and pi}
\end{center}
\end{figure}

As the time of each event $e$ is bounded by a global time interval $[lo,hi]$ (as shown in Fig.~\ref{F:space-time diagram}) and each local state $s$ is defined by its adjacent events $le(s)$ and $he(s)$, we can define the \textit{definite interval} and \textit{possible interval} of a local state. The definite interval $I_{def}(s)$ of local state $s$ is $$I_{def}(s)=[le(s).hi, he(s).lo]$$
indicating that the process is definitely in local state $s$ when the global time is in $I_{def}(s)$. Similarly, the possible interval $I_{pos}(s)$ of local state $s$ is $$I_{pos}(s)=[le(s).lo, he(s).hi]$$
indicating that the process is possibly in local state $s$ when the global time is in $I_{pos}(s)$. As in Fig.~\ref{F:di and pi}, $I_{def}(s^{(1)}_1)=[6,11]$ and $I_{pos}(s^{(1)}_1)=[4,13]$ with $\varepsilon=1$.

Notice that a CGS is a vector of local states from each process. We can further define the \textit{definite interval} and \textit{possible interval} of a CGS. The definite interval $I_{def}(\mathcal{C})$ of CGS $\mathcal{C}$ is the intersection of all the definite intervals of the constituent local states, i.e.,
$$I_{def}(\mathcal{C})=[\max\limits_{1\leq k \leq n}{I_{def}(\mathcal{C}[k]).lo}, \min\limits_{1\leq k \leq n}{I_{def}(\mathcal{C}[k]).hi}]$$
indicating that the system is definitely in CGS $\mathcal{C}$ when the global time is in $I_{def}(\mathcal{C})$ (when $I_{def}(\mathcal{C}).lo \leq I_{def}(\mathcal{C}).hi$). Similarly, the possible interval $I_{pos}(\mathcal{C})$ is the intersection of all the possible intervals of the constituent local states, i.e.,
$$I_{pos}(\mathcal{C})=[\max\limits_{1\leq k \leq n}{I_{pos}(\mathcal{C}[k]).lo}, \min\limits_{1\leq k \leq n}{I_{pos}(\mathcal{C}[k]).hi}]$$
indicating that the system is possibly in CGS $\mathcal{C}$ when the global time is in $I_{pos}(\mathcal{C})$. Take the CGS $\mathcal{C}_{11}$ in Fig.~\ref{F:Lattice} as an example, $I_{def}(\mathcal{C}_{11}) = [6,7]$ and $I_{pos}(\mathcal{C}_{11}) = [4,9]$, as shown in Fig.~\ref{F:di and pi}. Each CGS of the lattice in Fig.~\ref{F:Lattice} is equipped with time intervals, with the upper one indicating the definite interval and the lower one indicating the possible interval.

\subsubsection{Lattice as a Timed Automaton}\label{sec:lattice as a TA}
\begin{figure}[tbp]
\begin{center}
  \includegraphics[width=2.5in]{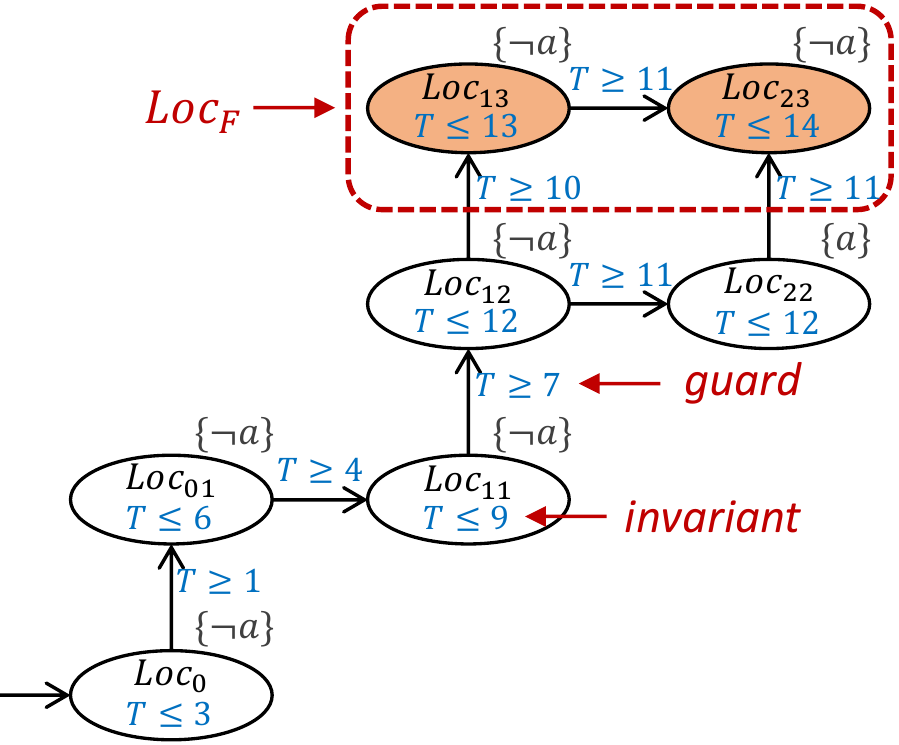}
  \caption{Timed automaton $TA(LAT)$}
  \label{F:timed automaton}
\end{center}
\end{figure}

The continuously-growing lattice equipped with time intervals can be characterized by a continuously-growing \textit{timed automaton} adapted from the classical timed automaton \cite{Alur94, Baier08}. The timed automaton corresponding to a currently-observed lattice is a tuple:
\begin{equation*}
  TA(LAT) = (Loc, T, \hookrightarrow, Loc_0, Inv, AP, L, Loc_F)
\end{equation*}
where
\begin{itemize}\setlength{\itemsep}{0pt}
    \item $Loc$ is a set of locations, i.e., the set of CGSs of $LAT$;
    \item $T$ is the global clock;
    \item $\hookrightarrow\ \subseteq Loc \times CC(T) \times Loc$ is a transition relation;
    \item $Loc_0 \in Loc$ is the initial location, i.e., the initial CGS $\mathcal{C}_0$ of $LAT$;
    \item $Inv: Loc \mapsto CC(T)$ is an invariant-assignment function;
    \item $AP$ is a finite set of all pre-defined CGS predicates;
    \item $L: Loc \mapsto 2^{AP}$ is a labeling function for the locations;
    \item $Loc_F \subseteq Loc$ is a finite set of accepting locations, i.e., the active surface CGSs $Act(LAT)$.
\end{itemize}

\noindent Here, $CC(T)$ denotes the set of clock constraints over $T$ in the form
\begin{equation*}
  g::= T < c~|~T\leq c~|~T > c~|~T \geq c~|~g \wedge g~(c\in \mathbb{N})
\end{equation*}

The transformation from the lattice in Fig.~\ref{F:Lattice} to a timed automaton is illustrated in Fig.~\ref{F:timed automaton}. The locations of the timed automaton correspond to the CGSs of the lattice. The \textit{invariant} of each location $\mathcal{C}$ is in the form $T\leq I_{pos}(\mathcal{C}).hi$, indicating the time that the system can stay at $\mathcal{C}$. The \textit{guard} of each transition to a location $\mathcal{C}$ is in the form $T \geq I_{pos}(\mathcal{C}).lo$, indicating when the transition can be taken.

The finite paths accepted by the timed automaton can be represented by a transition system $TS(TA(LAT))$ (or $TS(TA)$ for short) \cite{Baier08}. The states $S$ in $TS(TA)$ are defined as $S = \langle \mathcal{C}, t \rangle$, where $\mathcal{C}$ is the current location and $t$ is the current value of the clock $T$. The finite paths $Path_{fin}(TS(TA))$ of $TS(TA)$ are the active-surface-induced CGS sequences equipped with timestamps, in the form
\begin{eqnarray*}
    \pi = & \langle Seq[0], 0\rangle \stackrel{d_1}{\rightarrow} \langle Seq[0], d_1\rangle \stackrel{e}{\rightarrow} \langle Seq[1], d_1\rangle \stackrel{d_2}{\rightarrow} \hspace{0.65cm}\\
    & \langle Seq[1], d_1+d_2\rangle \stackrel{d_3}{\rightarrow} \cdots \stackrel{d_j}{\rightarrow} \langle Seq[i], d_1+\cdots+d_j\rangle
\end{eqnarray*}
with $Seq(\mathcal{C}_0, \mathcal{C}_i)\in Path(LAT)$, $Seq[i]\in Loc_F$, and $$I_{def}(Seq[i]).hi \leq d_1+\cdots+d_j \leq I_{pos}(Seq[i]).hi~\footnote{The lower bound of $d_1+\cdots+d_j$ can be tightly bounded by the value in Eq.~(\ref{eq:inf guard}) by replacing $\mathcal{C}$ with $Seq[i]$.}$$
We use $\pi[k]$ to denote the $(k+1)^{st}$ state of path $\pi$, use $\pi[k].\mathcal{C}$ to denote the location of $\pi[k]$, and use $\pi[k].t$ to denote the time value of $\pi[k]$. Notice that each path starts with the initial state $\langle Loc_0, 0\rangle$, and continuously grows as the lattice grows, to a potentially infinite size.

\section{Specification of Temporal Properties} \label{sec:Specification}

Our specification inherits the notions of branching time and metric time from TCTL \cite{Alur94}. It is a tailored subset of TCTL, which trades certain expressiveness for the efficiency of verification. The 3-valued semantics is adopted to cope with the verification over the currently-observed finite trace, in contrast to the traditional boolean semantics over the infinite trace of possible system execution. We first present the syntax and then discuss the 3-valued semantics.

\subsection{Syntax}

The syntax of our specification is defined as follows:
\begin{eqnarray*}
  \Phi &::=& \exists \varphi~|~\forall \varphi \\
  \varphi &::=& \Diamond ^J \phi~|~\Box ^J \phi \\
  \phi &::=& a~|~g~|~\phi \wedge \phi~|~\phi \vee \phi~|~\neg \phi~|~\phi \Rightarrow \phi
\end{eqnarray*}
where $a\in AP$, $g\in CC(T)$, and $J\subseteq\mathbb{R}_{\geq 0}$ is a time interval bounded by natural numbers \cite{Baier08}.

Our specification is a subset of TCTL without nested modal operators \cite{Behrmann04}. It can express numerous properties in our scenario. Informally, $\exists \Diamond ^J \phi$ means $\phi$ possibly holds during the interval $J$, $\exists \Box ^J \phi$ means $\Phi$ potentially always holds in $J$, $\forall \Diamond ^J \phi$ means $\phi$ always eventually holds in $J$, and $\forall \Box ^J \phi$ means $\phi$ invariantly holds in $J$. Notice that $\exists \Box ^J \phi = \neg \forall \Diamond ^J \neg \phi$ and $\forall \Box ^J \phi = \neg \exists \Diamond ^J \neg \phi$. For example, the property $C_1$ (i.e., ``\textit{all the robots should gather at the assembly point within 15 seconds}'') in Section~\ref{sec:Introduction} can be expressed as \textit{$\Phi_{C_1} = \forall \Diamond^{[0, 15]} a$ with $a$~=~``all the robots are at the assembly point''}.

\subsection{The 3-valued Semantics}\label{sec:Semantics}

We first discuss the classical boolean semantics over infinite paths, then discuss why the 3-valued semantics is inevitable for finite paths, and finally present the 3-valued semantics.

Given an infinite path $\widetilde{\pi}\in Path_{inf}(\widetilde{TS})$ of a transition system $\widetilde{TS}$ with infinite paths and a time $j \geq 0$, we can get the locations in the path at time $j$, denoted by
$$Loc(\widetilde{\pi}, j) =\{\widetilde{\pi}[k].\mathcal{C}~|~k\geq 0 \wedge \widetilde{\pi}[k].t\leq j \wedge \widetilde{\pi}[k+1].t\geq j\}$$
Note that we can easily check that whether a location $\mathcal{C}$ satisfies a predicate $\phi$, i.e., $\mathcal{C}\models \phi$.\footnote{The details are omitted here for brevity.} Thus, we can check whether an infinite path $\widetilde{\pi}$ satisfies a path formula $\varphi$:
\begin{eqnarray}
    \widetilde{\pi} \models \Diamond^J \phi & \textrm{iff} & \exists j\in J, \exists \mathcal{C}\in Loc(\widetilde{\pi}, j), \mathcal{C}\models \phi \\
    \widetilde{\pi} \models \Box^J \phi & \textrm{iff} & \forall j\in J, \forall \mathcal{C}\in Loc(\widetilde{\pi}, j), \mathcal{C}\models \phi
\end{eqnarray}
Then we can easily check whether a transition system $\widetilde{TS}$ with infinite paths satisfies a TCTL formula $\Phi$:
\begin{eqnarray}\label{eq:inf TS and phi}
\widetilde{TS} \models \exists \varphi & \textrm{iff } & \exists \widetilde{\pi}\in Path_{inf}(\widetilde{TS}), \widetilde{\pi}\models \varphi\\
\widetilde{TS} \models \forall \varphi & \textrm{iff } & \forall \widetilde{\pi}\in Path_{inf}(\widetilde{TS}), \widetilde{\pi}\models \varphi
\end{eqnarray}
The timed automaton $\widetilde{TA}$ of the transition system $\widetilde{TS}$ satisfies a TCTL formula $\Phi$ iff the transition system $\widetilde{TS}$ satisfies the formula $\Phi$ \cite{Baier08}, i.e.,
\begin{eqnarray}\label{eq:inf TA and Phi}
\widetilde{TA}\models \Phi & \textrm{iff } & \widetilde{TS}(\widetilde{TA})\models \Phi
\end{eqnarray}

However, the paths of the transition system corresponding to the lattice are finite, and finite paths may not be sufficient to either satisfy or falsify TCTL formulas \cite{Bauer11, Wei12}. For example, based on the classical boolean semantics of TCTL, the timed automaton in Fig.~\ref{F:timed automaton} does not satisfy the formula $\Phi_{C_1}$. The timed automaton grows as the lattice grows, and there may be a new successor location of $Loc_{13}$ and $Loc_{23}$ satisfying `$a$' (as in Fig.~\ref{F:extended true TA}), which may lead to the satisfaction of $\Phi_{C_1}$. That is to say, the classical semantics of TCTL does not provide intuitive and convenient support for this case of ``being inconclusive'' over the currently-observed finite trace. This case of being inconclusive may often appear when verifying TCTL formulas over the observed finite trace at runtime.

Discussions above motivate us to adopt the 3-valued semantics, i.e., providing a third value ``inconclusive'' for the case of being inconclusive \cite{Bauer11, Wei12}. We use the symbols `$\top$', `$\bot$', and `?' to denote ``\textit{true}'', ``\textit{false}'', and ``\textit{inconclusive}'', respectively. Let $\widetilde{\Pi}$ be the set of all the infinite timed paths with time non-decreasing. The semantics of whether a finite path $\pi$ satisfies a path formula $\varphi$ is defined as follows:
\begin{eqnarray}\label{eq:path and varphi}
[\pi \models \varphi] = \left\{ \begin{array}{ll}
\top & \textrm{if } \forall\sigma, \pi\sigma\in\widetilde{\Pi}, \pi\sigma\models\varphi\\
\bot & \textrm{if } \forall\sigma, \pi\sigma\in\widetilde{\Pi}, \pi\sigma\not\models\varphi\\
? & \textrm{otherwise.}
\end{array} \right.
\end{eqnarray}
The semantics of our specification is defined as follows:
\begin{eqnarray}
[TS(TA) \models \exists \varphi] \hspace{2.2in}\nonumber\\ =\left\{ \begin{array}{ll}
\top & \textrm{if } \exists \pi\in Path_{fin}(TS(TA)), [\pi\models \varphi] = \top\\
\bot & \textrm{if } \forall \pi\in Path_{fin}(TS(TA)), [\pi\models \varphi] = \bot\\
? & \textrm{otherwise.}
\end{array} \right.
\end{eqnarray}
\vspace{-0.15in}
\begin{eqnarray}
[TS(TA) \models \forall \varphi] \hspace{2.2in}\nonumber\\ =\left\{ \begin{array}{ll}
\top & \textrm{if } \forall \pi\in Path_{fin}(TS(TA)), [\pi\models \varphi] = \top\\
\bot & \textrm{if } \exists \pi\in Path_{fin}(TS(TA)), [\pi\models \varphi] = \bot\\
? & \textrm{otherwise.}
\end{array} \right.
\end{eqnarray}
\vspace{-0.15in}
\begin{eqnarray}\label{eq:TA and Phi}
[TA\models \Phi] = \left\{ \begin{array}{ll}
\top & \textrm{iff } [TS(TA)\models \Phi] = \top\\
\bot & \textrm{iff } [TS(TA)\models \Phi] = \bot\\
? & \textrm{otherwise.}
\end{array} \right.
\end{eqnarray}

\section{Verification of the Specified Property at Runtime} \label{sec:Detection}

In this section, we discuss the verification of the specified property at runtime. Each time a new local state of some process is sent to $P_{che}$, $P_{che}$ first incrementally updates the new active surface, constructs the corresponding timed automaton, and then checks the property $\Phi$ over the timed automaton. The checking of $\Phi$ over the timed automaton is achieved by checking $\Phi$ over two special extended timed automata with infinite paths.

We first discuss the incremental maintenance of the active surface and the corresponding timed automaton, and then discuss the verification of the specified property.

\subsection{Maintenance of the Timed Automaton}

$P_{che}$ continuously collects the execution trace from the processes, and maintains the active surface (not the whole lattice of system snapshots) at runtime. The maintenance of the active surface is incremental in that new CGSs can grow from the active surface. With the notion of the active surface, evolution of the lattice can be viewed as discarding the ``old'' nodes in the active surface and obtaining the ``new'' ones incrementally at runtime \cite{Yang13}. The worst-case number of active surface CGSs is in $O(np^{n-1})$, where $p$ is the upper bound of the number of local states of each process, and $n$ is the number of processes. (Note that the worst-case number of CGSs of the whole lattice is $O(p^n)$.) Please refer to our previous work \cite{Yang13} for more detailed discussions on the runtime maintenance algorithm of the active surface.

Based on the incremental maintenance of the active surface, the timed automaton corresponding to the lattice can also be incrementally constructed. Notice that each location of the timed automaton corresponds to one CGS of the lattice. The successor CGSs of the inactive CGSs (e.g., the white CGSs in Fig.~\ref{F:Lattice}) have already been discovered, and the corresponding part of the timed automaton (e.g., the white locations in Fig.~\ref{F:timed automaton}) also has been completely constructed. Thus, the timed automaton can be incrementally constructed based on the runtime maintenance of the active surface. Whenever a new active CGS is added to the active surface, a new location corresponding to the new CGS is added to the timed automaton.

\subsection{Verification of the Specified Property}\label{sec:Detection over TA}

Though our specification is a tailored subset of TCTL, we cannot directly apply the standard TCTL model checking algorithms on the timed automaton $TA$ due to the 3-valued semantics dedicated for finite paths of $TS(TA)$. According to the 3-valued semantics in Eq.~(\ref{eq:path and varphi}), we should append each finite path $\pi$ with all possible infinite suffixes $\sigma$ and check whether the infinite paths $\pi\sigma\in \widetilde{\Pi}$ satisfy the path formula $\varphi$. Rather than appending the finite path with all possible infinite suffixes, we define two special types of infinite suffixes $\sigma_\top$ and $\sigma_\bot$ with $\forall k, \sigma_\top[k].\mathcal{C} \models \phi$ and $\sigma_\bot[k].\mathcal{C}\not\models \phi$, to ease the verification. Then, Eq.~(\ref{eq:path and varphi}) can be rewritten as follows:
\begin{eqnarray}
[\pi \models \varphi] = \left\{ \begin{array}{ll}
\top & \textrm{if } \pi\sigma_\top\models\varphi \wedge \pi\sigma_\bot\models\varphi\\
\bot & \textrm{if } \pi\sigma_\top\not\models\varphi \wedge \pi\sigma_\bot\not\models\varphi\\
? & \textrm{otherwise.}
\end{array} \right.
\end{eqnarray}
\begin{figure}[tbp]
\begin{center}
  \includegraphics[width=2.5in]{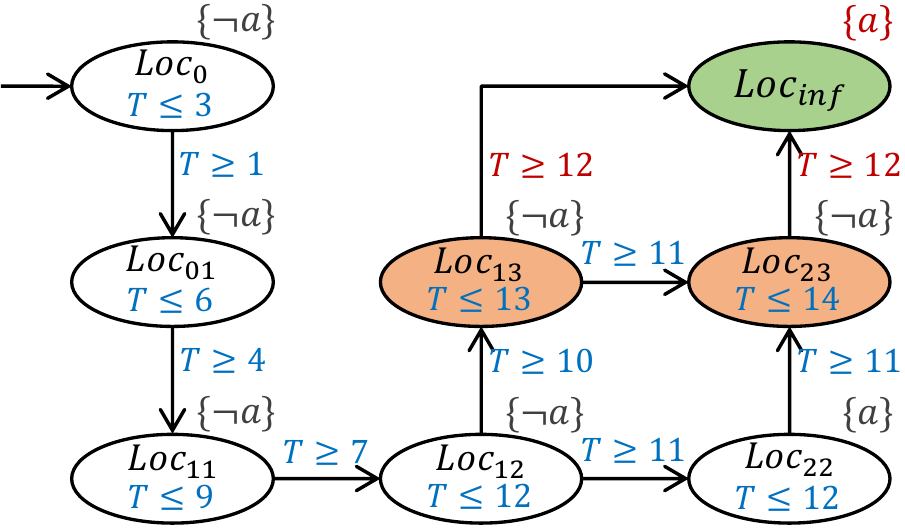}
  \caption{Extended timed automaton $\widetilde{TA}_\top$}
  \label{F:extended true TA}
\end{center}
\end{figure}
\begin{figure}[tbp]
\begin{center}
  \includegraphics[width=2.5in]{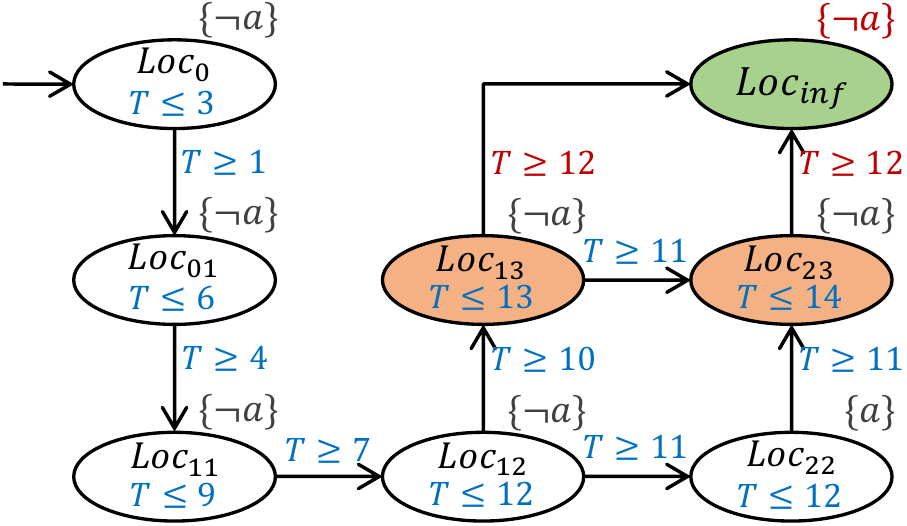}
  \caption{Extended timed automaton $\widetilde{TA}_\bot$}
  \label{F:extended false TA}
\end{center}
\end{figure}

Based on the semantics above, we can extend each path $\pi\in Path_{fin}(TS(TA))$ with the two types of infinite suffixes $\sigma_\top$ and $\sigma_\bot$. The extension of $\sigma_\top$ is achieved by adding an extra location $Loc_{inf}$ with $L(Loc_{inf})=\{\phi\}$ to the timed automaton, and adding transitions from each of the locations $\mathcal{C}\in Loc_F$ to $Loc_{inf}$ with the guard
\begin{eqnarray}\label{eq:inf guard}
T\geq \min\limits_{i\in \{k|\mathcal{C}[k] = \mathcal{G}_{max}[k]\}} {\max(I_{pos}(\mathcal{C}).lo, I_{def}(\mathcal{C}[i]).hi)}
\end{eqnarray}
The guard can be understood by assuming that, in dimension $i$ of the CGS $\mathcal{C}$, a successor CGS $\mathcal{C}'$ is coming. The guard from $\mathcal{C}$ to $\mathcal{C}'$ is $T\geq I_{pos}(\mathcal{C}').lo = \max (I_{pos}(\mathcal{C}).lo, I_{pos}(\mathcal{C}'[i]).lo)$ with $I_{pos}(\mathcal{C}'[i]).lo = I_{def}(\mathcal{C}[i]).hi$. The extension of $\sigma_\bot$ is in the same way except that $L(Loc_{inf})=\{\neg\phi\}$. After that, we can get two extended timed automata $\widetilde{TA}_\top$ and $\widetilde{TA}_\bot$ with infinite paths. The extensions to the timed automaton of Fig.~\ref{F:timed automaton} are shown in Fig.~\ref{F:extended true TA} and Fig.~\ref{F:extended false TA}. The location $Loc_{inf}$ in Fig.~\ref{F:extended true TA} is labeled with $\{a\}$ and the location $Loc_{inf}$ in Fig.~\ref{F:extended false TA} is labeled with $\{\neg a\}$. In Fig.~\ref{F:extended true TA}, based on Eq.~(\ref{eq:inf guard}), the guard of the transition from $Loc_{13}$ to $Loc_{inf}$ is $T\geq \max(I_{pos}(\mathcal{C}_{13}).lo, I_{def}(\mathcal{C}_{13}[2]).hi)$. Notice that $I_{pos}(\mathcal{C}_{13}).lo = 10$ (as shown in Fig.~\ref{F:Lattice}), and $I_{def}(\mathcal{C}_{13}[2]).hi = I_{def}(s^{(2)}_{3}).hi = 12$ (as shown in Fig.~\ref{F:space-time diagram}). Thus, the guard of the transition from $Loc_{13}$ to $Loc_{inf}$ is $T\geq 12$.

Based on Eq.~(\ref{eq:inf TS and phi})-(\ref{eq:inf TA and Phi}), we can check whether $\widetilde{TA}_\top\models \Phi$ and $\widetilde{TA}_\bot\models \Phi$. Then the semantics of our specification in Eq.~(\ref{eq:TA and Phi}) can be rewritten as follows:
\begin{eqnarray}\label{eq:TA and inf TA}
[TA \models \Phi] = \left\{ \begin{array}{ll}
\top & \textrm{if } [\widetilde{TA}_\top\models \Phi]=[\widetilde{TA}_\bot\models \Phi]=\top\\
\bot & \textrm{if } [\widetilde{TA}_\top\models \Phi]=[\widetilde{TA}_\bot\models \Phi]=\bot\\
? & \textrm{otherwise.}
\end{array} \right.
\end{eqnarray}
Consequently, the verification boils down to the verification of $\widetilde{TA}_\top\models \Phi$ and $\widetilde{TA}_\bot\models \Phi$. Based on the two timed automata in Fig.~\ref{F:extended true TA} and Fig.~\ref{F:extended false TA}, we can know that $[TA \models \Phi_{C_1}] =\ ?$, i.e., the result is ``inconclusive''.

The verification of the formula $\Phi$ on the timed automaton $\widetilde{TA}$ is achieved by checking the derived CTL formula $\hat{\Phi}$ on $\widetilde{TA}$ \cite{Baier08}. The time interval $J$ in the formula $\Phi$ is eliminated by adding equivalent clock constraints into the $\phi$ part. We can transform the TCTL formula $\Phi$ into a CTL formula $\hat{\Phi}$ as follows\footnote{As we do not allow nested modal operators in the specification and the timed automaton has no clock resets, there is no need for introducing a fresh clock to measure the elapse of time, as in \cite{Baier08}.}:
\begin{eqnarray*}
\Phi = \exists \varphi & \textrm{then} & \hat{\Phi} = \exists \hat{\varphi}\\
\Phi = \forall \varphi & \textrm{then} & \hat{\Phi} = \forall \hat{\varphi}\\
\varphi = \Diamond ^J \phi & \textrm{then} & \hat{\varphi} = \Diamond (T\in J \wedge \phi)\\
\varphi = \Box ^J \phi & \textrm{then} & \hat{\varphi} = \Box (T\in J \Rightarrow \phi)
\end{eqnarray*}
For example, the derived CTL formula of $\Phi_{C_1} = \forall \Diamond^{[0, 15]} a$ is $\hat{\Phi}_{C_1} = \forall \Diamond (T\leq 15) \wedge a)$.

The equivalence of the satisfaction is ensured as follows:
\begin{eqnarray}\label{eq:Phi and CTL Phi}
\widetilde{TA} \models \Phi & \textrm{iff} & \widetilde{TA} \models_{CTL} \hat{\Phi}
\end{eqnarray}
The proof is straightforward and omitted here \cite{Baier08}. Thus, the verification is finally reduced to the verification of $\widetilde{TA}_\top\models_{CTL} \hat\Phi$ and $\widetilde{TA}_\bot\models_{CTL} \hat\Phi$, i.e., checking a CTL formula (without nested modal operators) on a classical timed automaton (with only one clock), which can be efficiently achieved by numerous optimization algorithms \cite{Baier08, Dill89, Laroussinie04, Wang04}. The skeleton of the verification algorithm is shown in Algorithm~\ref{A:Checker}.
\begin{algorithm}[tbp]
\SetAlgoVlined
\textbf{Upon} initialization\\
\hspace{0.4cm}get property $\Phi$, and transfer $\Phi$ into CTL formula $\hat{\Phi}$\;
\textbf{Upon} receiving local state $s^{(k)}_i$ from $P^{(k)}$\\
\hspace{0.4cm}construct $Act(LAT)$ with $s^{(k)}_i$ incrementally\;
\hspace{0.4cm}construct $TA(LAT)$ incrementally\;
\hspace{0.4cm}extend $TA(LAT)$ into $\widetilde{TA}_\top$ and $\widetilde{TA}_\bot$\;
\hspace{0.4cm}check $\widetilde{TA}_\top \models_{CTL} \hat{\Phi}$ and $\widetilde{TA}_\bot \models_{CTL} \hat{\Phi}$\;\label{line:check}
\hspace{0.4cm}check $TA \models \Phi$ according to Eq.~(\ref{eq:TA and inf TA})-(\ref{eq:Phi and CTL Phi})\;
\caption{Verification algorithm on $P_{che}$\label{A:Checker}}
\end{algorithm}

\section{Experiments} \label{sec:Performance measurements}

In this section, we first describe the implementation of \textsf{PARO}, and then discuss the performance evaluation. The effectiveness of \textsf{PARO} is demonstrated through a case study of a realistic mobile robot gathering scenario based on our RobotCar project. Details of the case study can be found in Appendix \ref{sec:Case Study}.

\subsection{Implementation}

We implement \textsf{PARO} on the open-source middleware we developed - {\it Middleware Infrastructure for Predicate detection in Asynchronous environments} (MIPA) \cite{MIPA}. Based on MIPA, we specify properties in TCTL formulas (e.g., the formula $\Phi_{C_1}$) to the middleware using an XML schema. The devices (e.g., mobile robots) register themselves to the middleware (abstracted as processes), and continuously send the trace with local timestamps to the checker processes on the middleware. Checker processes are implemented as third-party services on the middleware, in charge of collecting related traces and verifying the formulas. The verification of $\widetilde{TA}_\top \models_{CTL} \hat{\Phi}$ and $\widetilde{TA}_\bot \models_{CTL} \hat{\Phi}$ (line 7 of Algorithm~\ref{A:Checker}) is achieved by incrementally generating XML descriptions for the timed automata and automatically invoking UPPAAL \cite{Behrmann04}, which is a toolbox for verification of real-time systems. Each time a formula is verified ``true'', ``false'', or ``inconclusive'', the middleware will notify the users.

\subsection{Performance Evaluation}\label{sec:Performance Evaluation}

In this section, we conduct simulations of the robot gathering scenario to evaluate the performance of \textsf{PARO} under different settings of key environmental factors. We first describe the experiment setup and then discuss the evaluation results.

We let the robots collect sensing data every second. We generate the sensing data using the Poisson distribution. Specifically, the average time of local activities (where the local predicate is true) on the robots is 10 s, and the average interval between the activities (where the local predicate is false) is 5 s. The number of the sensing data items on each robot is up to 2,400. The lifetime of the experiments is up to 40 mins. The experiments are conducted on a PC running Windows 8.1 (x64) and Java version 1.7 with an Intel Core i5-2400 Quad-Core Processor (3.10 GHz) and 8 GB of memory.

In the experiments, we check the formula $\forall \Diamond^{[0, 15000]} (LP_1\wedge \cdots \wedge LP_n)$\footnote{The local predicate $LP_i = (R_i.front \leq 600~mm \wedge R_i.left \leq 400~mm)$, indicating robot $R_i$ is at the assembly point.} and tune two key environmental factors - the number of processes (i.e., the mobile robots) $n$ and the clock difference bound $\varepsilon$ - to evaluate the five performance metrics $S_M$, $S_U$, $T_M$, $T_U$, and $|Loc|$. $S_M$ denotes the average memory cost of MIPA (mainly for the construction of the active surface of the lattice and the timed automaton), $S_U$ denotes the average memory cost of UPPAAL, $T_M$ denotes the average time cost for the construction of the active surface and the timed automaton by MIPA (line 4-6 of Algorithm~\ref{A:Checker}), $T_U$ denotes the average time cost for the verification by UPPAAL (line 7-8 of Algorithm~\ref{A:Checker}), and $|Loc|$ denotes the size of the timed automaton $TA$ when the experiment stops. $S_M + S_U$ indicates the average of the total memory cost, and $T_M + T_U$ indicates the average of the total latency.

\subsubsection{Effects of Tuning the Number of Processes}

In this experiment, we study how the number of processes $n$ affects the performance of \textsf{PARO}. We fix the clock difference bound $\varepsilon$ to 200 ms, and tune $n$ from 2 to 30.
\begin{figure}[tbp]
\begin{center}
  \includegraphics[width=2.5in]{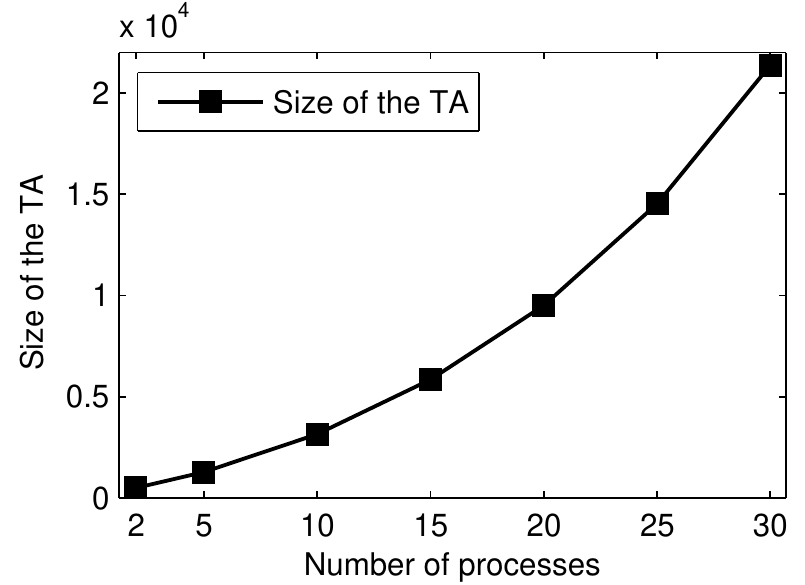}
  \caption{The size of the $TA$ vs. the number of processes ($\varepsilon$ = 200 ms)}
  \label{F:size TA np}
\end{center}
\end{figure}

As shown in Fig.~\ref{F:size TA np}, the increase of $n$ leads to fast increase in the size of the $TA$ (from 484 to 21,354). Compared to the size of $LAT$ under the asynchronous model in our previous work \cite{Yang13}, the size of the lattice (i.e., $|Loc|$) in this work is rather smaller. To a certain extent, the lattice structure turns out to be applicable on the partially synchronous model, although it is exponential on the asynchronous model. The reason is that, we have made use of the existing synchrony of the system and efficiently restricted the size of the lattice, comparing to previous work under the asynchronous model \cite{Schwarz94, Yang13}.
\begin{figure}[tbp]
\begin{center}
  \includegraphics[width=2.5in]{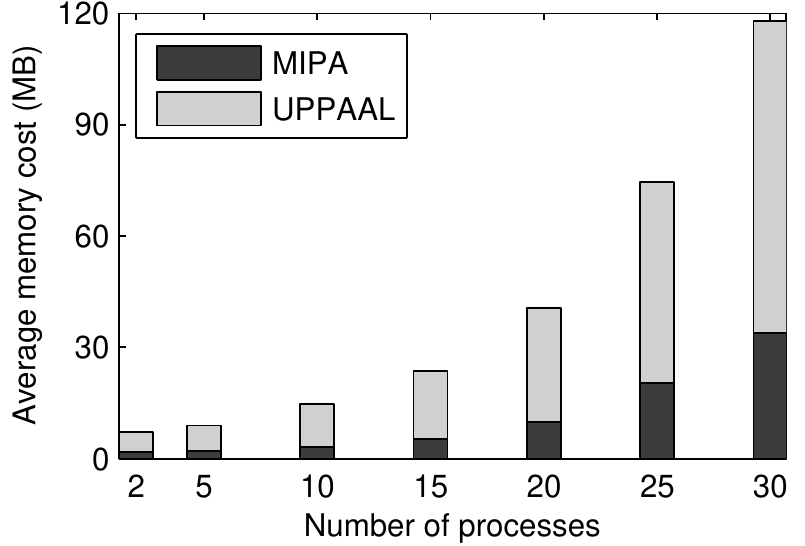}
  \caption{Average memory cost vs. the number of processes ($\varepsilon$ = 200 ms)}
  \label{F:memory vs np}
\end{center}
\end{figure}
\begin{figure}[t]
\begin{center}
  \includegraphics[width=2.5in]{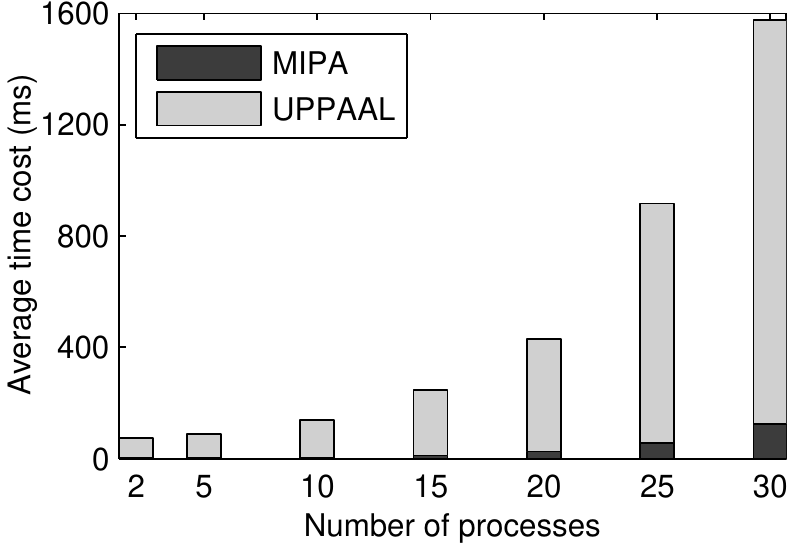}
  \caption{Average time cost vs. the number of processes ($\varepsilon$ = 200 ms)}
  \label{F:time vs np}
\end{center}
\end{figure}

As shown in Fig.~\ref{F:memory vs np}, the average memory cost of MIPA $S_M$ increases quickly as $n$ increases, while $S_U$ (the average memory cost of UPPAAL) increases a little faster than $S_M$. As we tune $n$ from 2 to 30, $S_M$ and $S_U$ increase from 1.8 MB to 34 MB and from 5.3 MB to 84 MB, respectively. That is to say, the memory cost of verification is larger than that of the construction of the active surface and the timed automaton. Moreover, the increase of $S_M + S_U$ is roughly in accordance with the increase of the size of the $TA$ in Fig.~\ref{F:size TA np}. The reason is that our TCTL formulas have no nested modal operators, thus can be space-efficiently verified over the $TA$.

As for the time cost, the average time cost of MIPA $T_M$ increases slowly from 1.8 ms to 125 ms as $n$ increases from 2 to 30, as shown in Fig.~\ref{F:time vs np}. However, the average time cost of UPPAAL $T_U$ increases quickly from 72 ms to 1,450 ms. Most of the time (over 90\%) is spent on the verification. The average of the total latency is acceptable. Furthermore, the increase of $T_U$ is roughly in accordance with the increase of the size of the $TA$ in Fig.~\ref{F:size TA np}. That is to say, \textsf{PARO} is relatively time-efficient.

\subsubsection{Effects of Tuning the Clock Difference Bound}

In this experiment, we study how the clock difference bound $\varepsilon$ affects the performance of \textsf{PARO}. We fix the number of processes $n$ to 10, and tune $\varepsilon$ from 10 ms to 1 s.

\begin{figure}[tbp]
\begin{center}
  \includegraphics[width=2.5in]{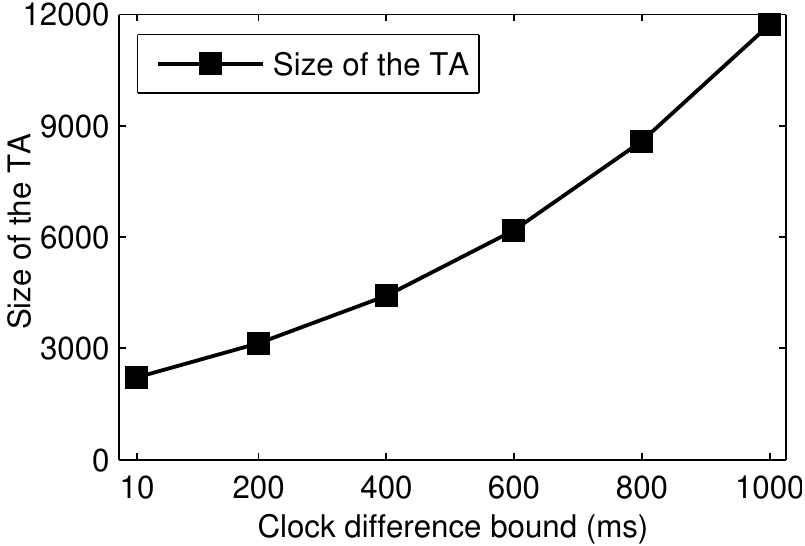}
  \caption{The size of the $TA$ vs. the clock difference bound $\varepsilon$ ($n$ = 10)}
  \label{F:size TA e}
\end{center}
\end{figure}

As shown in Fig.~\ref{F:size TA e}, the increase of $\varepsilon$ leads to quick increase in the size of the $TA$ (from 2,220 to 11,722). This is because the increase of $\varepsilon$ leads to more possible interleavings of events, thus leads to more possible system snapshots. Compared to Fig.~\ref{F:size TA np}, the increase of the size of the $TA$ caused by $\varepsilon$ is slower than that caused by $n$, i.e., the number of processes $n$ has greater impact on the size of the $TA$ than the clock difference bound $\varepsilon$. As the bound of clock difference is often small in realistic systems, the number of processes $n$ becomes the most important environmental factor.

\begin{figure}[tbp]
\begin{center}
  \includegraphics[width=2.5in]{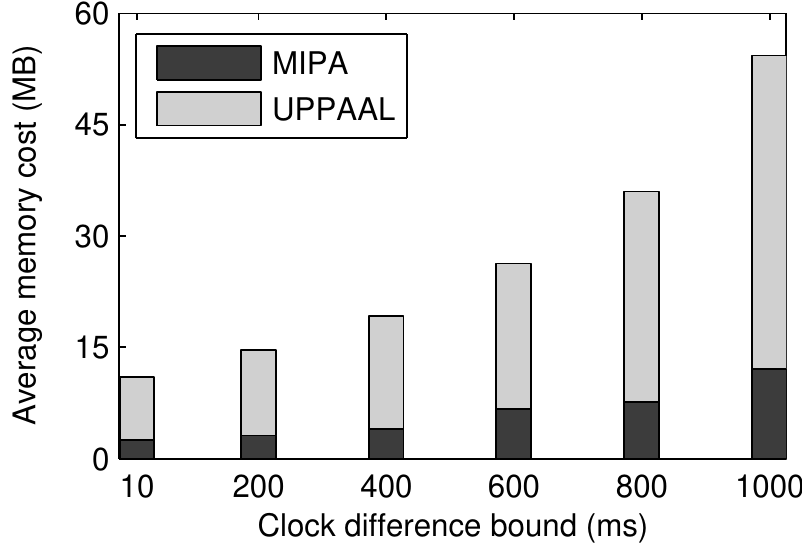}
  \caption{Average memory cost vs. the clock difference bound $\varepsilon$ ($n$ = 10)}
  \label{F:memory vs e}
\end{center}
\end{figure}

As shown in Fig.~\ref{F:memory vs e}, the average memory cost of MIPA $S_M$ increases slowly as $\varepsilon$ increases, while $S_U$ (the average memory cost of UPPAAL) increases a little faster than $S_M$. As we tune $\varepsilon$ from 10 ms to 1,000 ms, $S_M$ and $S_U$ increase from 2.6 MB to 12 MB and from 8.5 MB to 42 MB, respectively. The memory cost of verification by UPPAAL is larger than that of the construction of the active surface and the timed automaton by MIPA, as that in Fig.~\ref{F:memory vs np}. Moreover, the increase of $S_M + S_U$ is roughly in accordance with the increase of the size of the $TA$ in Fig.~\ref{F:size TA e}. The reason is the same as that for the number of processes. Consequently, \textsf{PARO} is relatively space-efficient.
\begin{figure}[tbp]
\begin{center}
  \includegraphics[width=2.5in]{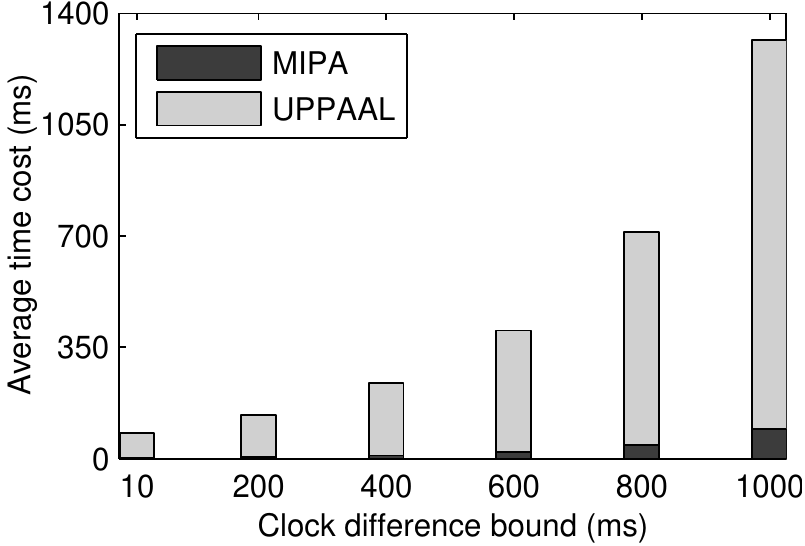}
  \caption{Average time cost vs. the clock difference bound $\varepsilon$ ($n$ = 10)}
  \label{F:time vs e}
\end{center}
\end{figure}

As for the time cost, the average time cost of MIPA $T_M$ increases slowly from 2.2 ms to 93 ms as $\varepsilon$ increases from 10 ms to 1,000 ms, as shown in Fig.~\ref{F:time vs e}. However, the average time cost of UPPAAL $T_U$ increases quickly from 78 ms to 1,222 ms. Most of the time (over 90\%) is spent on the verification. The average of the total latency is acceptable. Notice that although $T_U$ increases quickly, it is roughly in accordance with the increase of the size of the $TA$ in Fig.~\ref{F:size TA e}. That is to say, \textsf{PARO} is relatively time-efficient.

\section{Related Work} \label{sec:Related work}

Our work can be posed against two areas of related work: model checking of real-time systems and detection of global predicates over distributed computations.

In the area of model checking of real-time systems, the existing work is mostly studied and developed in the framework of Alur and Dill's Timed Automata, and TCTL gains extensive research since then \cite{Baier08, Alur93, Alur94}. Most work focuses on finding efficient algorithms for timed automata verification \cite{Baier08, Dill89, Laroussinie04, Wang04}. These optimization algorithms are generally orthogonal to \textsf{PARO}. Although \textsf{PARO} shares many similarities with model checking over timed automata, there are important differences. First, the complete model (e.g., a series of timed automata) of the system is mandatory in model checking over timed automata \cite{Baier08, Alur93, Alur94}. In contrast, as an external observer of an already-running system, our work is applicable to ``black box'' systems with no system model at hand. Model checking deals with infinite traces of all possible executions, whereas our work deals with the currently-observed finite trace of one concrete execution, by modeling the finite trace as a continuously-growing timed automaton. The temporal logics of model checking, such as TCTL, are interpreted over infinite traces, and the checking cost is often prohibitive \cite{Baier08}. However, to trade the expressiveness for the checking cost, our specification is a tailored subset of TCTL without nested modal operators, as in \cite{Behrmann04}. Moreover, unlike the classical semantics of TCTL over infinite traces, we adopt 3-valued semantics for our specification over the currently-observed finite trace to cope with the case of ``being inconclusive''.

In detection of global predicates over distributed computations (aka ``runtime verification''), existing work can be categorized by the timing model. Based on the synchronous model, predicates can be detected over a single total order of events \cite{Bauer11, Kshemkalyani07, Xu05}. Bauer et al.~\cite{Bauer11} detect 3-valued semantics of LTL and TLTL over a sequence of timed events. Our 3-valued semantics of TCTL is partially inspired by this work \cite{Bauer11}. Kshemkalyani \cite{Kshemkalyani07} detects predicates concerning the relationships of time intervals over event streams. Xu et al.~\cite{Xu05} detect first-order logic based formulas over collected contextual activities. However, the actual system is imperfectly synchronized in our scenario, which makes the work above inapplicable. Based on the asynchronous model, predicates are detected over multiple possible total orders of events consistent with the `happen-before' relation resulting from message passing \cite{Cooper91, Babaoglu93, Schwarz94, Garg96}, as in our previous work \cite{Huang09, Huang12, Yang13, Wei12}. We detect conjunctive predicates in \cite{Huang09, Huang12}, regular expression predicates in \cite{Yang13}, and 3-valued CTL predicates in \cite{Wei12}. However, the asynchronous model is overly-pessimistic in that the existing synchrony of the system is completely abandoned, and the detection over the asynchronous model is usually expensive \cite{Yang13, Wei12}. Based on the partially synchronous model, predicates can be detected at a relatively lower cost \cite{Marzullo91, Mayo95, Stoller00, Duggirala12}. Marzullo et al.~\cite{Marzullo91}, Mayo et al.~\cite{Mayo95}, and Stoller \cite{Stoller00} detect global predicates without timing constraints in partially synchronous systems. Duggirala et al.~\cite{Duggirala12} detect whether there exists a real-time $t$, when bounded executions of the system that correspond to the trace satisfy a given global predicate. In the work \cite{Duggirala12}, the complete system model as a timed input/output automaton is mandatory. In contrast, our \textsf{PARO} works over the trace with no system model at hand. We model the observed trace as a continuously-growing timed automaton, and check a subset of TCTL formulas with 3-valued semantics over the timed automaton.

\section{Conclusion and Future Work} \label{sec:Conclusion}

In this work, we propose the \textsf{PARO} framework for formal specification and runtime verification of properties with metric-time constraints over the trace of a partially synchronous system of mobile robots. The \textsf{PARO} framework consists of three essential parts: 1) modeling of the trace of system execution; 2) specification of temporal properties; 3) verification of the specified property at runtime.

In our future work, we need to design an optimized incremental algorithm for the detection of $\widetilde{TA}_\top\models_{CTL} \hat\Phi$ and $\widetilde{TA}_\bot\models_{CTL} \hat\Phi$, since the automaton $TA(LAT)$ is incrementally constructed as the active surface of the lattice evolves. We also need to dig into the specific type of timed automaton corresponding to the timed trace, and study how to detect the full TCTL over the timed automaton while preserving a relatively low checking cost. In this work, we investigate partially synchronous systems with the processes synchronizing their clocks with an external source clock. In our future work, we need to investigate runtime verification of properties in partially synchronous systems with the processes synchronizing their clocks internally. A more comprehensive experimental evaluation is also necessary.

\section*{Acknowledgements}

This work is supported by the National 973 Program of China (2015CB352202), the National Science Foundation of China (61272047, 91318301, 61321491), and the Program A for Outstanding PhD candidate of Nanjing University. We are grateful to Lei Bu's insightful comments and Chao Fang's implementations on the mobile robots.

\bibliographystyle{IEEEtran}
\bibliography{IEEEabrv,TR042015}

\newpage
\appendix

\subsection{Case Study}\label{sec:Case Study}

The case study of a realistic robot gathering scenario is conducted based on our RobotCar project\footnote{The RobotCar project: http://cs.nju.edu.cn/yuhuang/robotcar.htm.}, to demonstrate the effectiveness of \textsf{PARO}. We first describe the scenario and then discuss the effectiveness.

\subsubsection{The Robot Gathering Scenario}

\begin{figure}[tbp]
\begin{center}
  \includegraphics[width=3.3in]{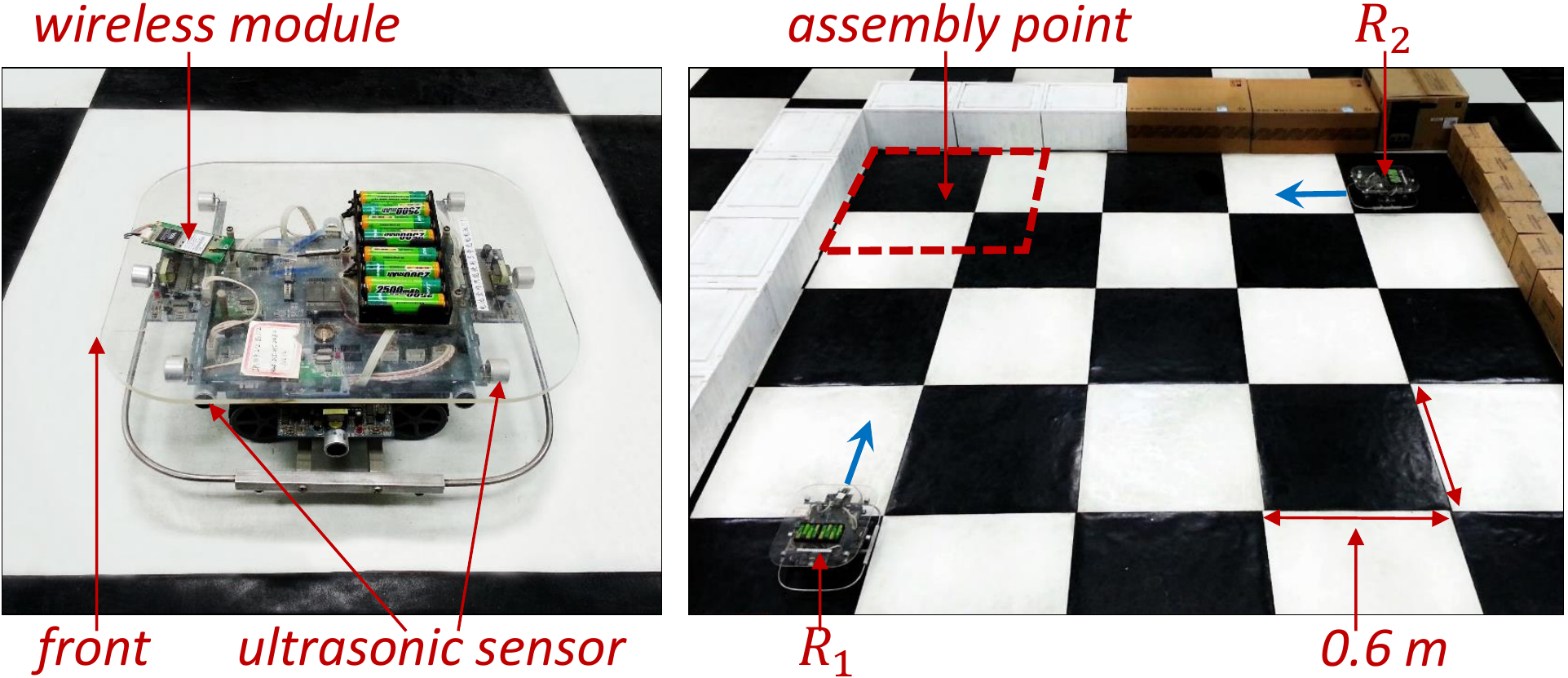}
  \caption{The mobile robot and the scenario}
  \label{F:case study}
\end{center}
\end{figure}

In this scenario, two mobile robots are designed to move along the wall and coordinate to meet at the assembly point of a $3~m \times 3~m$ room (for further collaboration), as shown in Fig.~\ref{F:case study}. Each robot is equipped with four ultrasonic ranging sensors for the front, back, left, and right directions, as well as a wireless module for communication. The robots synchronize their clocks with a timing server in the same room. The actual difference $\varepsilon$ between the clocks of the robots and the timing server is bounded by 10 ms. The robots collect the readings from the four sensors every second and label the readings with local timestamps. The robots start from two corners of the room, move at the speed of about 0.12 m/s, and adjust their routes according to the readings of the four sensors (to keep a constant distance from the wall) and the status (e.g., the location and the speed) of the other robot.

In the standard way of thinking about the computation, the system execution is regarded as a totally-ordered progression of the system state. From the view of users of the robots, the two mobile robots are expected to gather at the assembly point within 15 seconds as specified in property $C_1$ (in Section \ref{sec:Introduction}). However, this synchronous model of time does not work in a partially synchronous system. Developers of the mobile robot system need to change to the notion of partially synchronous time. Then they can observe and analyze the execution of the mobile robots under the guidance of the \textsf{PARO} framework, as detailed below.

\subsubsection{Effectiveness}

Based on \textsf{PARO}, temporal properties with different real-time constraints can be conveniently specified and verified over the system execution trace at runtime. The property is first specified to MIPA. As the robots move, they send the timed trace (i.e., sensing data with local timestamps) to MIPA (running on a dedicated server in the room), and MIPA verifies the TCTL formulas at runtime.

We use the conjunction of two local predicates to express that the two robots arrive at the assembly point, i.e., $a = LP_1\wedge LP_2$. The local predicate $LP_1 = (R_1.front \leq 600~mm \wedge R_1.left \leq 400~mm)$, indicating $R_1$ is at the assembly point. Similarly, the local predicate $LP_2 = (R_2.front \leq 600~mm \wedge R_2.right \leq 400~mm)$, indicating $R_2$ is at the assembly point.

As for one single timed path of the system execution, we only have to check whether the path satisfies $\Diamond^{[0, 15000]} a$.\footnote{In this scenario, we take 1 ms as the unit of time.} However, due to the intrinsic asynchrony in our scenario, we can get multiple possible timed paths of the system state. Thus we have to employ the modal operators ($\forall$ and $\exists$) in TCTL to cope with the uncertainty resulting from multiple possible system executions. Specifically, we specify a pair of TCTL formulas to MIPA:
$$\Phi_{C_1} = \forall \Diamond^{[0, 15000]} a \textrm{ and } \Phi'_{C_1} = \exists \Diamond^{[0, 15000]} a$$
There are three different situations according to the satisfaction of $\Diamond^{[0, 15000]} a$ under different modalities:
\begin{enumerate}\setlength{\itemsep}{0pt}
    \item All possible timed paths of the system state satisfy $\Diamond^{[0, 15000]} a$, i.e., $\Phi_{C_1}$ is true. In this case, the real-time constraint is definitely satisfied in all possible executions;
    \item Some (but not all) timed paths of the system state satisfy $\Diamond^{[0, 15000]} a$, i.e., $\Phi_{C_1}$ is false and $\Phi'_{C_1}$ is true. This shows that although some paths satisfy the property, the property is possible to be violated. It indicates that the program logics of the robots have potential bugs, since the property cannot be guaranteed in all possible executions. Put it in another way, even though the property is satisfied in the current execution, it is possible that the property is violated, e.g., in the next execution;
    \item  None of the timed paths of the system state satisfies $\Diamond^{[0, 15000]} a$, i.e., $\Phi'_{C_1}$ is false. Since the property is violated in all possible executions, we can infer that, either the specified property is beyond the capacity of the mobile robots, or there are severe flaws in the system implementation.
\end{enumerate}

In our case study, we start the robots, check the formulas by MIPA, and find that $\Phi_{C_1}$ is checked false and $\Phi'_{C_1}$ is checked true. It indicates a potential violation of the property. Thus, if the user requires that the robots always gather in time, the developers should revise the program logics of the robots, e.g., by explicitly handling the uncertainty result from the asynchrony.

We further conduct two experiments to verify another two pairs of TCTL formulas. The first pair is dedicated for checking the property $C_2=$ ``whether the robots can gather within 14,000 ms'': $\Phi_{C_2}=\forall \Diamond^{[0, 14000]} a$ and $\Phi'_{C_2}=\exists \Diamond^{[0, 14000]} a$. Formulas $\Phi_{C_2}$ and $\Phi'_{C_2}$ are checked false, which means that the robots cannot gather within 14 s. Another pair is dedicated for checking the property $C_3=$ ``whether the robots can gather within 16,000 ms'': $\Phi_{C_3}=\forall \Diamond^{[0, 16000]} a$ and $\Phi'_{C_3}=\exists \Diamond^{[0, 16000]} a$. Formulas $\Phi_{C_3}$ and $\Phi'_{C_3}$ are checked true, which means that the robots can definitely gather within 16 s.

In summary, by specifying properties with different real-time requirements, and verifying the properties under different modalities based on \textsf{PARO}, we can gain a deep insight into the runtime behavior of the mobile robot system.

Furthermore, our 3-valued semantics is well motivated in the experiments. During the experiments, when the observed finite execution trace is not sufficient to satisfy or falsify the formulas (e.g., when the time of observation is less than 14 s), we are encountered with the case of ``being inconclusive'', while applying the classical boolean semantics may lead to false positive or false negative.

We further describe the actual performance of \textsf{PARO} in this case study. We evaluate the metrics $S_M$, $S_U$, $T_M$, and $T_U$ defined in Section \ref{sec:Performance Evaluation}. The performance of the case study is shown in Table~\ref{T:case study}. We can see that the average of the total memory cost $S_M + S_U$ is less than 5 MB while the average of the total latency $T_M + T_U$ is small (less than 50 ms). This proves that \textsf{PARO} can verify the formulas in time and is feasible in this kind of realistic scenarios.
\begin{table}[tbp]
    \caption{Experimental results of the case study}
    \label{T:case study}
    \centering
    \begin{tabular}{cc|cccc}\hline
        $n$ & $\varepsilon$ (ms) & $S_{M}$ (MB) & $S_{U}$ (MB) & $T_{M}$ (ms) & $T_{U}$ (ms) \\ \hline\hline
        2 & 10 & 1.8 & 2.2 & 0.5 & 41.0 \\ \hline
    \end{tabular}
\end{table}

\end{document}